# Slider on a Driven Substrate:
# Markovian Competition Between Two Limiting Attractors

Juleon M. Schins and Diego Maza Ozcoidi

Departamento de Física y Matemática Aplicada, Facultad de Ciencias, Universidad de Navarra, Pamplona, Spain

*Abstract*

*We present experimental data of the motion of a rimmed checker's piece on a polished horizontal tray, in two specific conditions: with the substrate harmonically shaken at 20 Hz, and with a static substrate. The latter experiment immediately yields the dynamic friction coefficient. The harmonic experimental results are very sensitive on the static friction coefficient, which is not even a percent higher than the dynamic one. Due to the low harmonic acceleration of the driver, the static friction has enormous influence on the slider's motion. We modelled the slider's motion using a discrete Markovian progression model, which at every discrete time-step chooses between a sticking attempt and a return to the non-sticking trajectory. The present model does not take into account the driver's acceleration. In the final section, we explain the physical origin of walk-off, in case of periodic but asymmetric driving.*

Version: August 9[th], 2018

# 1. Introduction

The classic laws of sliding friction were discovered by Leonardo da Vinci in 1493, a pioneer in tribology, but the laws documented in his notebooks were not published [DaV] and became only recently known as being identical to those of Guillaume Amontons published in 1699 [Amo], and that of Coulomb published in 1821 [Cou]. In 1750 Leonhard Euler derived the angle of repose of a weight on an inclined plane and first distinguished between static and dynamic friction. Ever since, increasingly sophistic models of friction have been published, each successful for the particular domain they were developed for. Today many tribologists believe that long-range attractive isotropic Van der Waals forces cause dynamic friction, and shorter-range attractive anisotropic Van der Waals forces, static friction. The only Van der Waals forces that survive in gas or liquid phases are London dispersion forces [Lon], the isotropic quantum-electrodynamic attraction between fluctuating charges on neutral molecules without permanent multipoles. In the solid phase, an analogous elimination of anisotropic forces occurs, as long as the slider is moving on its substrate. When the slider stalls, after some settling (or "repose") time, the slider may microscopically reorient and find a position in which also anisotropic Van der Waals forces add to the dynamic binding. Hence, interlocking asperities only play a tribologic role in unpolished systems.

On a macroscopic scale, the usually extremely short-range Van der Waals forces survive in the form of Hamaker's [Ham] elaboration of Fritz London's 1937 description of dispersion [Lon]. When a sphere is much closer to a plane than its radius, with a surface-to-surface separation of *d*, Hamaker's attraction is proportional to the radius, and inversely proportional to the squared distance. This dispersive force has the longest range, after the interaction between two charged objects. Although its range is identical to the interaction between a charged object and a permanent dipole, its strength is many orders of magnitude weaker.

It would be questionable to state that to date static friction is a univocal term. Too many different models exist, describing too many different specific circumstances [refs], for such a statement to be true. This paper presents a tractable macroscopic sliding model for polished surfaces, with a novel approach to the treatment of static friction. Here the adjective "tractable" signifies an *a posteriori* surface characterization by means of a few macroscopic fit parameters, as opposed to an *a priori* molecular mapping of the two surfaces involved. Although the latter is, to date, technically feasible for nano-surfaces, it is still years out of reach for macroscopic surfaces. Moreover, even if one went into the painstaking data accumulation of the molecular structure of two macroscopic surfaces, the experiments would require nano-precision initial conditions. The question remains, what such an experiment could possibly teach about static friction.

Our approach simply assumes the existence of nanoscale irregularities on both surfaces, and treats them in a probabilistic way. The probabilistic element is therefore neither of the most fundamental kind, like quantum indeterminacy, but of a huge-number, practically unknowable but fundamentally knowable kind, like Boltzmann's thermodynamics: it trades in all-comprehensive knowledge of a single deterministic system for for knowledge of averages in the context of a theory with a fundamental lack of all-comprehensive knowledge. In spite of the model's saturation with fit parameters, we hope to take up Feynman's challenge, who claimed, "with dry metals it is very hard to show any difference" (between the dynamic and static friction coefficients) [Fey].

We performed two preliminary experiments to achieve an equally preliminary model describing the response of a slider to a harmonically moving substrate. The first experiment records the deceleration of the slider moving with an initial velocity on an otherwise static substrate. This experiment allows a quite precise determination of the dynamic friction coefficient. The second experiment is the harmonically driven slider. The harmonic data are a box of surprises, which we tried to explain one by one.

The main idea of our model is Markovian propagation within the bounds of two limiting "attractor curves". One limit is that in which the slider is permanently stuck to the harmonically moving substrate: in this case, the slider and driver trajectories are identical, which always occurs in the limit of low accelerations. The other limit is the "stick-free Van der Waals curve": an idealized curve the slider would follow in case the velocity and acceleration differences between slider and substrate are high enough not to allow any static (short-range) Van der Waals bonding to occur. In this case, the dynamic and static friction coefficients are numerically equal. In Figs. 2.2 and 3.1 these "limiting attractor curves" are depicted blue and green. The Markovian walk of the slider occurs largely between the two limiting attractors, in such a way, that for nearly coinciding velocity and acceleration of slider and substrate, there is a strong tendency for the slider to stick to the substrate.

Whereas the driver curve simply depends on the the experimental settings of amplitude and frequency (in case of harmonic motion of the substrate), the stick-free curve has to be "construed" via a trial-and-error procedure. This results in a somewhat erratic-appearing organization of the chapters. Although the Markovian progression law does contain the physics of linear relaxation to the stick-free curve, the enhanced sticking probability whenever the two attractor-curves cross requires fit parameters. In first instance, one knows neither the fit parameters of the Markovian sticking probability, nor those defining the stick-free limiting curve; worst of all, parameter changes in one domain influence the other. The fit parameter optimization therefore starts by guessing all fit parameters, and then, with an eye on the experimental result, optimizing them in order to get a best possible overlap between measurement and model prediction. This procedure is trial-and-error because one continuously alternates between the parameter domain of the stick-free curve and that of the attraction mechanism.

The last section of this paper applies the resulting model with its now fixed set of parameters, to the case of asymmetric driving. Without the need of any adaptable parameters, we find that our model explains the origin of walk-off.

## 2. Harmonic data

The slider is a rimmed cylinder (1 cm length, 2.5 cm diameter) placed like a checker's piece onto a tray. A fluorescent reference is glued onto the tray; the cylinder is equally provided with a fluorescent disk. The tray moves harmonically in a horizontal plane, under diffuse white illumination. The blue curves in Fig. 2.1 represent 18 periods of measured reference *positions*, and the orange curves the co-measured cylinder positions; that is, the two positions are determined from a single camera shot. The appendix explains the smoothing details (see Eq. A1 and Fig. A1).

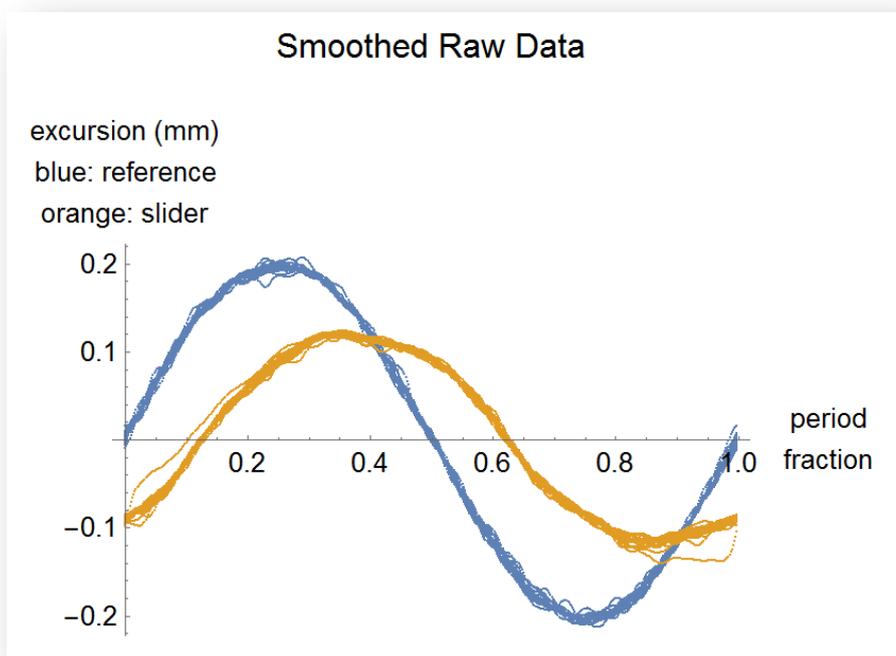

*Figure 2.1. Measured Positions of Reference (blue) and Slider (orange). The traces represent 18 periods of measurement, superposed within a single period, after removal of digitizing errors of the fast camera. The horizontal scale counts the number of periods; such that unity period corresponds to 50 ms, the period of a 20 Hz harmonic motion. Note that the data are laboratory-referenced. This implies that a sign-inverted response represents the highest relative velocity with respect to the moving substrate, and a driver-superposed response implies stalling.*

The camera is set at 10 shots per millisecond, which means that a single period contains 500 data points. Fig. 2.2 illustrates the periodic driver and slider *velocities*. With respect to a sinusoidal shape, there is a clear deformation of the slider's velocity, which looks more like a triangle. Strong oscillations locally affect both slider and driver traces.

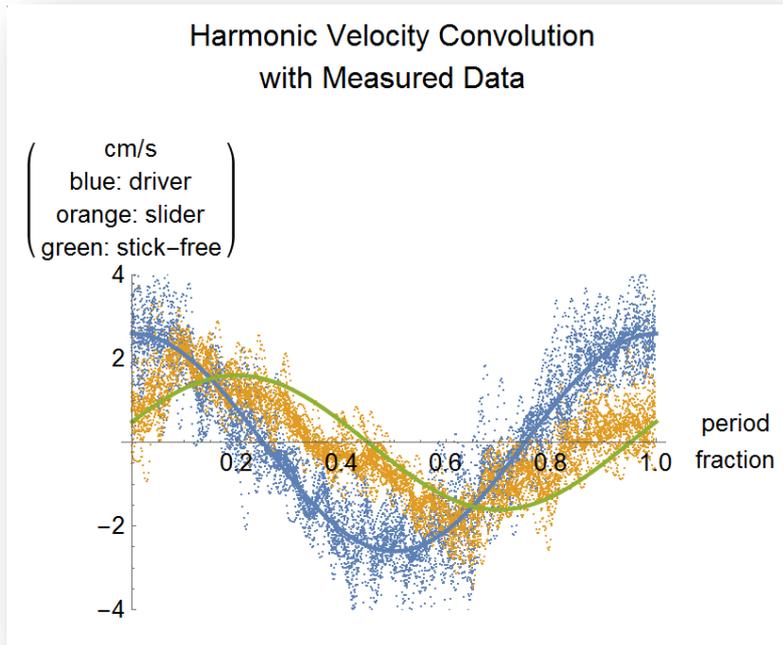

*Figure 2.2: Measured Velocities of Driver and Slider.* *Removal of the highest frequency components of the Fourier spectrum of the traces grants the absence of digitizing errors. Consequently, all oscillations superposed on the fundamental are real velocity paths, not due to digitizing errors. The green line is drawn for future purposes.*

Although the driver's velocity approaches a pure cosine (disregarding high-frequency oscillations), the slider's velocity is closer to a triangle than a sinusoid. Instead of defining and solving a non-linear equation, we first calculate the ideal motion of a slider that has only "non-sticking Van-der-Waals contact" with its ideal substrate, and only busy ourselves with the "sticking Van-der-Waals contact" in Sections 4 and 7. In a macroscopic sense, the term "stick-free" signifies the numerical equality of the dynamic and static friction coefficients.

Throughout the paper, we use the words "driver", "reference", and "substrate" for a single object, the first one in the context of forces, the second one in the context of measured signals, and the third one in the context of friction with the slider. Moreover, throughout the paper, velocities are laboratory-referenced, unless specifically indicated.

# 3. Linear Kernel Convolution Based on Momentum Transfer

Quite generally, a mathematical convolution describes an object's reaction to a non-instantaneous force. In other terms, the object has a "memory" of past forces, and its reaction is an integral over past forces. The general connection between a "kick-response" of a once-hit slider $K_{kick}(t)$, the slider's acceleration $a_{slide}(t)$, and its response to a driving force $F(t)$ is

[3.1]
$$\begin{cases} m\ddot{x}_{conv}(t) = \int_{-\infty}^{\infty} ds\, K_{kick}(s) F_{drive}(t-s) \\ m\dot{x}_{conv}(t) = \int_{-\infty}^{\infty} ds\, K_{kick}(s) p_{drive}(t-s) \\ mx_{conv}(t) = \int_{-\infty}^{\infty} ds\, K_{kick}(s) mx_{drive}(t-s) \end{cases}$$

In all three cases, the kernel $K_{kick}$ is identical. It is a normalized quantity in frequency units, defined as the ratio of the kick-momentum $p_{kick}(t) = mv_{kick}(t)$ –with $m$ the slider's mass– and its absolute time integral:

[3.2] $$K_{kick}(t) \equiv \frac{p_{kick}(t)}{\left|\int_{-\infty}^{\infty} ds\, p_{kick}(s)\right|} = \frac{v_{kick}(t)}{\left|\int_{-\infty}^{\infty} ds\, v_{kick}(s)\right|}$$

We define the driver's harmonic velocity as $v_{drive}(t) = v_{d0} \cos \omega_0 t$, with radial frequency $\omega_0 = 2\pi \cdot 20 Hz$. The displacement amplitude is $x_{d0} = 0.21\,mm$, as can be directly read off Fig. 2.1. This corresponds to a velocity amplitude of $v_{d0} \equiv \omega_0 x_{d0} = 2.6\,cm\,s^{-1}$. The normalized kernel has the linear shape:

[3.3] $$K_{kick}(t\,|\,t_{mx}) \equiv \frac{2}{t_{mx}}(1 - \frac{t}{t_{mx}})\theta(t)\theta(t_{mx} - t)$$

The Heaviside function is defined as usual: $\theta(t) = \begin{cases} 1 & \text{for } t \geq 0 \\ 0 & \text{for } t < 0 \end{cases}$. The normalization grants that

[3.4] $$\int_{-\infty}^{\infty} dt\, K_{kick}(t\,|\,t_{mx}) \equiv 1$$

This property of the kernel, combined with its finite duration $t_{mx}$, add an additional property to the kernel: when $t_{mx} \downarrow 0$, the kernel turns into a delta function. In that limit, the convolution response of the slider falls on top of the driver: the slider is "permanently stuck" to the substrate. The latter solution applies close to a black hole, due to the extremely high value of gravitational constant. As the velocity reads

[3.5] $$\left|v_{kick}(t\,|\,t_{mx})\right| \equiv \left|v_{d0}\right|(1 - \frac{t}{t_{mx}})\theta(t)\theta(t_{mx} - t) = (\left|v_{d0}\right| - \mu_{dyn} gt)\theta(t)\theta(t_{mx} - t)$$

the time

$$[3.6] \quad t_{mx} \equiv \frac{|v_{d0}|}{\mu_{dyn} g}$$

is nothing but an alias for the friction, with $g = 9.8 ms^2$ the earth's gravity constant, and $\mu_{dyn}$ the dynamic friction coefficient. The "stick-free" or "Van der Waals" velocity response therefore reads

$$[3.7] \quad \begin{cases} v_{stick-free}(t \,|\, t_{mx}) = \frac{2}{t_{mx}} \int_0^{t_{mx}} ds\, (1 - \frac{s}{t_{mx}}) v_{d0} \cos \omega_0 (t-s) = \\ = 2 v_{d0} \frac{\cos \omega_0 t - \cos \omega_0 (t - t_{mx}) + \omega_0 t_{mx} \sin \omega_0 t}{(\omega_0 t_{mx})^2} = \frac{2 v_{d0}}{(\omega_0 t_{mx})^2} A_{harm} \cos(\omega_0 t - \phi_{harm}) \end{cases}$$

with the definitions

$$[3.8] \quad \begin{cases} A_{harm} \equiv \sqrt{A_c^2 + A_s^2} \\ \tan \phi_{harm} \equiv \frac{A_s}{A_c} \end{cases} \quad \begin{cases} A_c \equiv 1 - \cos \omega_0 t_{mx} \\ A_s \equiv \omega_0 t_{mx} - \sin \omega_0 t_{mx} \end{cases}$$

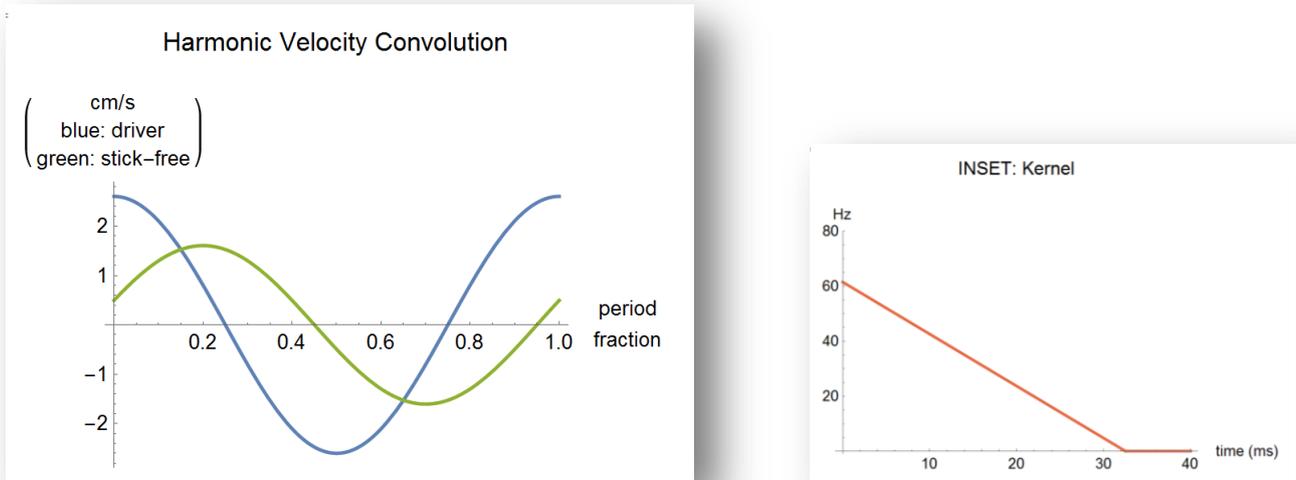

*Figure 3.1: Harmonic Velocity Convolution. The stick-free response (green curve) is a convolution of the driver velocity (blue curve) with the unit-surface kernel shown in the inset (red curve).*

The stick-free harmonic velocity (green curve in Fig. 3.1) is a delayed cosine, given by Eq. 7.1. We anticipate a paradox of the here presented kernel: its free-flight time is three times longer than expected. This paradox will be discussed in Section 6.

# 4. The Reach of a Harmonic Convolution

In this section, we study what combination of amplitude and phase are possible using an ordinary harmonic convolution, by plotting Eqs. 3.7 as a function of the free-flight time, in units of driving period.

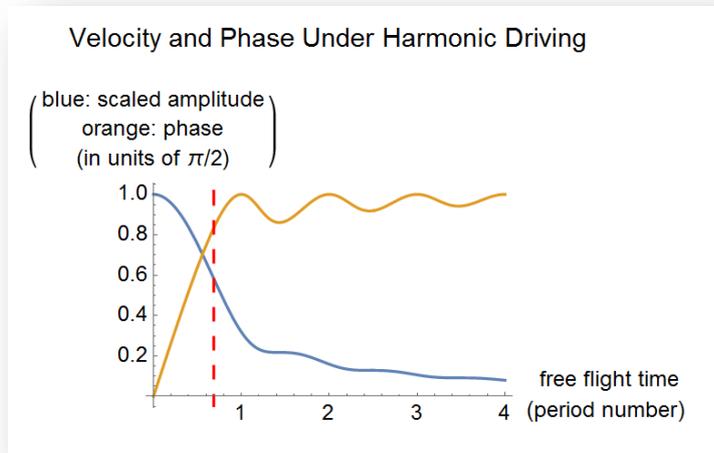

*Figure 4.1: Behavior of the Slider's Scaled Amplitude and Phase as a Function of Free Flight Time. The free flight time (see the inset in Fig. 3.1) is in units of period fraction. The blue curve represents the slider's amplitude divided by the driver's: it is always below unity due to energy conservation. The orange curve represents the phase delay of slider with respect to driver, in units of π/2. The dashed red line represents the here used free flight time.*

Fig. 4.1 illustrates the reach of a convolution: not all combinations of amplitude and phase are possible. In our specific case (the green curve in Fig. 3.1), the response amplitude is 60% of the driver's amplitude, and the phase delay is $0.4\pi$, as indicated by the red dashed line in Fig 3.2, and it is realized at a kernel duration of 65% period fraction. To reach a sine-like response, one might use a kernel-duration of 100%, but that automatically reduces the convolution-to-driver amplitude ratio to 30%. Note that, due to Eq. 3.6, the amplitude-phase solution does not change, as long as the ratio of $v_{d0}$ (driver amplitude) and $\mu_{dyn}$ (dynamic friction) remains constant.

# 5. Noise

Fig. 2.1 illustrates 18 superposed traces, both for driver and slider *position*. Without additional information, it is difficult to know which pixel belongs to which of the 18 series. Hence, we order the 18 traces and Fourier transform each of them; we then apply the "noise filter" (described in the appendix), which removes the signal frequencies and the frequencies in the neighborhood of the Nyquist frequency (thus removing the digitizing errors); we next back-transform the resulting data to (real) time-dependent noise, and average 18 periods. Fig. A1.2 (see appendix) shows the result of this procedure. Fig 5.1 shows a low-frequency pass version.

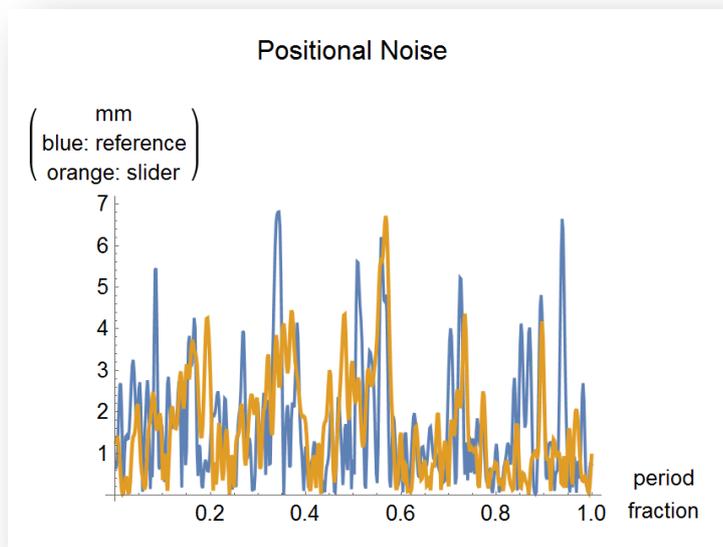

*Figure 5.1: Low-Frequency Passed Noise on the Position Data. The vertical scale follows from the equations used to calculate the noise (see appendix). Note that the horizontal axis represents time, not frequency.*

We now have all ingredients to attribute the major peaks in the noise spectrum. For that purpose, we plot all information gathered in the previous sections into a single Fig. 5.2. Somewhat tentative are the following noise attributions:

1. The driver noise peak at period fraction 9% marks the avalanche-like sticking of sliders that previously adhered to the stick-free curve.
2. At period fraction 17% the stick-free and driver velocities overlap: the absence of large noise signifies a strong reduction of un-sticking attempts.
3. The noise peak on the slider at period fraction 19% characterizes successful unsticking from the tray.

4. The noise peak on the driver at period fraction 26% indicates collective sticking attempts.
5. The noise on both slider and driver in the interval 33% to 39% characterizes simultaneous sticking and un-sticking attempts. It is like an interval of insecurity: far away from both attraction curves.
6. The lack of noise at period fraction 42% coincides with the crossing of the acceleration curves (the derivatives of the blue and green sinusoids in Fig. A2).
7. The heavier high-frequency oscillations superposed on the driver velocity represent a periodic phenomenon (as opposed to noise); most importantly, the superposed oscillations do not always coincide with those on the slider motion. We do not understand the nature of these features.

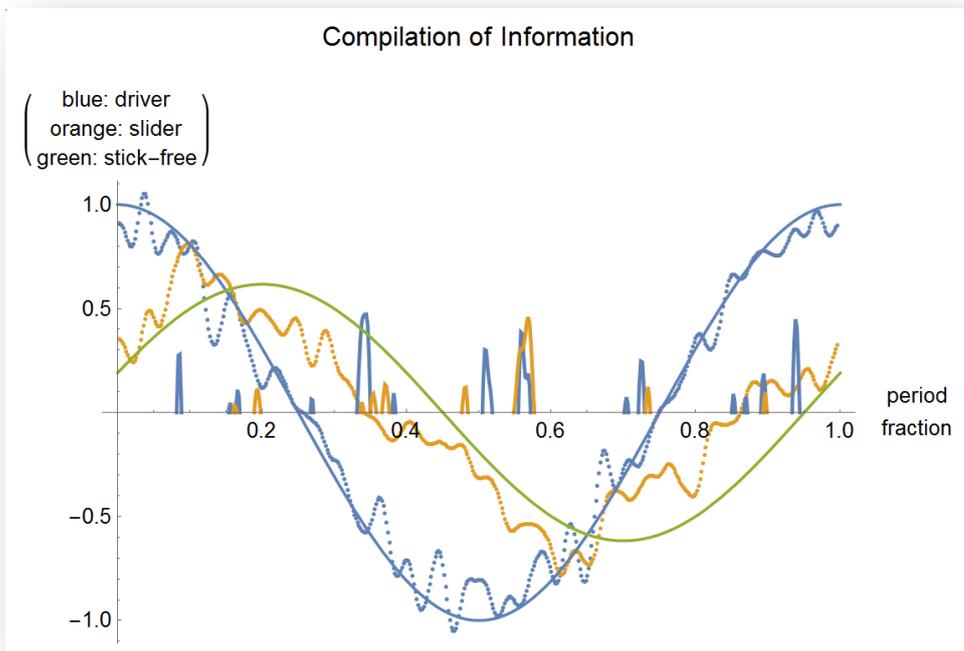

***Figure 5.2: Combination Plot of All Data.*** *This plot resumes (i) the 18-period averaged experimental velocity data (dots), (ii) the idealized attractor sinusoids (driver and stick-free velocities), and (iii) the noise peaks (thick), taken from Fig. 5.1. Clearly, the slider data always meander between the two attractor curves.*

# 6. Experimental Kick-Response Halted by Sticking

Kick responses are much simpler to measure than the harmonic responses. All one does is place the rimmed cylinder on a flat, static, well-polished glass substrate, and capture its deceleration (using the high-speed camera, still set at 10 shots per millisecond) after an initial kick. Fig. 6.1 shows the measured position kick-response data.

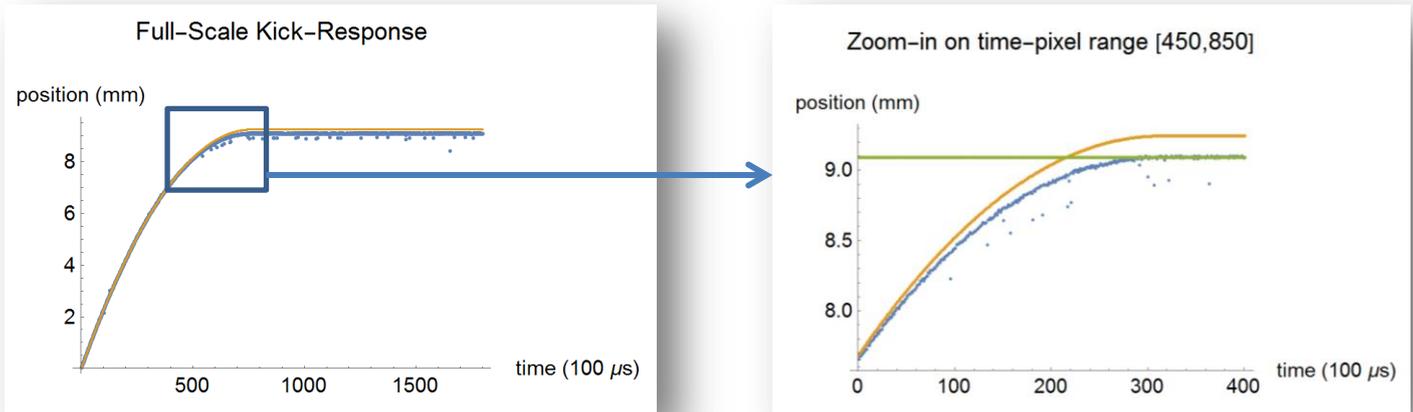

*Figure 6.1: Measured Positional Kick-Response. The LHS plot shows all the experimental data (blue dots), as well as a quadratic polynomial fit over the time-pixel interval [0,500] (orange parabola, the natural stick-free trajectory). The RHS plot shows a zoom-in (with shifted time-pixel scale) of the LHS plot starting at its pixel 450. After the parabola reaches its apex, the slider stalls (pixels 762 through 1800 in the LHS plot). The total time scale (1800 time pixels) spans 180 milliseconds: clearly visible is the slider's stalling at about the same time where the parabola reaches its apex. The horizontal green line represents the saturation level of the data.*

The orange curve represents the stick-free trajectory of a slider with numerically equal dynamic and static coefficients. The second time derivative of the experimental position data yields a dynamic friction coefficient of $\mu_{dyn} g \equiv -\ddot{x}_{parabola} \Rightarrow \mu_{dyn} = 32\%$. It is a quite precise estimation due to the 40 milliseconds long overlap time of data and stick-free parabola (Fig. 6.1, LHS plot).

Now we have all ingredients to discuss the paradox of the factor-3-longer-than-expected free-flight time of the kernel (inset in Fig. 3.1). Filling in the numbers, Eq. 3.6 predicts $\mu_{dyn} \approx 10\%$. This number is certainly wrong, given the enormous range over which the measured kick data show that $\mu_{dyn} = 32\%$. So what is the matter? Well, from Fig. 2.1 one knows that the maximum displacement amplitude of the driver is $x_{d0} = 0.21\, mm$. Therefore (see zoom-in of Fig 6.1), the typical excursion of the driver reaches only from the stalling point at 9.1 *mm* to 8.9 *mm*! Evidently, an eventual quadratic fit of that tiny part of the kick-deceleration curve would yield a much lower friction coefficient. This number has very little in common with the measured dynamic friction coefficient, heavily affected as it is by the static friction coefficient. *Stating the same idea in other words, the velocity or acceleration amplitude of the driver is at least a factor 3 too low to have the slider probe even a tiny fraction of the dynamic friction coefficient.*

We dedicate the final part of this Section to the dynamics of a single sticking event. In order to bring those dynamics better to the light, we plot the difference between the actual position of the slider and the expected stick-free trajectory.

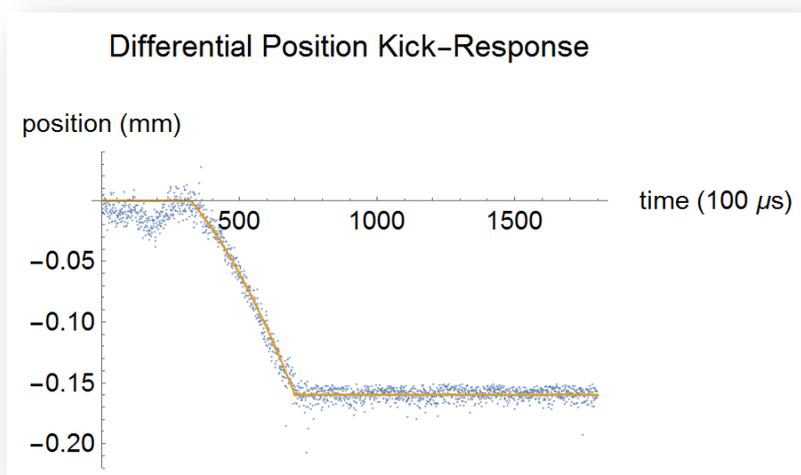

*Figure 6.2: Dynamics of a Sticking Attempt. The blue dots represent the difference position data with respect to the parabola. The orange curve is a piecewise quadratic fit over the pixel interval [320,662].*

Fig. 6.2 shows it took the slider 34 milliseconds to realize the jump from its stick-free trajectory to sticking. Possibly, more attempts preceded the successful jump at pixel 320. Since no external force operated on the slider throughout the measurement, "sticking" equals "definitive stalling".

# 7. Discrete-Markovian Walk Model

The main ingredients of the discrete-Markovian walk model are

(i) the existence of two "attractor curves": on one hand the driver's or "permanently stuck slider curve", and on the other, the "stick-free Van der Waals curve" (in Figs. 2.2 and 3.1 these attractor curves are depicted blue and green);

(ii) the probabilistic nature of the slider's decision, at each time step, to move to either attractor curve.

As mentioned at length in the introduction, the probability aspect is not of a fundamental kind, but of a "circumstantial" kind: a macroscopic experiment is always subject to unknown mesoscopic structures that play up in different ways according to the exact initial conditions and the nanoscale landscape of the two surfaces involved. We here present a crude model, which neglects acceleration in general. The model is Markovian, discrete in both time and velocity. Consider the velocity in stick-free equilibrium to be the convolution of the driver velocity with the kick-response of $\omega_0 t_{mx} = 2\pi \times 0.53$:

[7.1] $\quad v_{stick-free}(t) = 60\% \, v_{d0} \cos(\omega_0 t - 0.4\pi)$

Define the difference velocity with respect to the slider velocity

[7.2] $\quad \Delta v_{sfs}(t) \equiv v_{stick-free}(t) - v_{slide}(t)$

According to the fundamental equation of motion, this difference decays linearly in time as

[7.3] $\quad \left| \Delta v_{sfs}(t_0 + \Delta t) \right| = \left| \Delta v_{sfs}(t_0) \right| - \mu_{dyn} g \Delta t$

unless a sticking attempt occurs before reaching the stick-free equilibrium ($\Delta v_{sfs} = 0$). Clearly, Eq. 7.3 defines the slider attraction to the stick-free limit. The opposite tendency is due to Van-der-Waals sticking attempts, characterized by sudden breaks on the relative velocity of slider with respect to driver. If these two tendencies are strong enough, the slider path remains confined between the two limiting curves. Short of any information concerning the exact form of the sticking force, we assume only the most necessary:

(i) a sticking attempt reduces the slider's velocity with respect to the driver's by a $\Delta\mu$-dependent amount, thereby approaching the slider to the reference curve;

(ii) the probability for a sticking attempt is strictly Markovian (memoryless);

(iii) the sticking attempts are distributed according to an ad-hoc probability function depending on both $v_{drive}$ and $v_{stick-free}$.

The discretized time pixels are $t_j \equiv j\delta\tau$ with $\delta\tau$ the Markovian time step. The discretized velocity states are $v_{slide} \equiv n\delta v$ with $n \in \mathbb{Z}$ and $\delta v$ the Markovian velocity step. In the following equations, we will make use of three typical discrete functions:

(i) the function "rd", which rounds off its real argument to the nearest integer;
(ii) the function "boole", which takes the value 1 if its argument is true, and 0 if its argument is false;
(iii) the function "ran", which yields a random number between zero and one.

Consider the following *Ansatz* for a Markovian velocity path in terms of the discretized velocity, that is, for the signed number $n$:

$$[7.4] \quad n_{j+1}(n_j \,|\, a) = n_j + P_a(n_j) \left\{ \begin{array}{l} (n_{j+1}^{[drive]} - n_j)\delta_{n_j^{[drive]}, n_j} + \\ +(1 - \delta_{n_j^{[drive]}, n_j}) M_d^{-1} \,\mathrm{sgn}[n_{j+1}^{[drive]} - n_j] \end{array} \right\} + M_{sf}^{-1} \overline{P}_a(n_j) \,\mathrm{sgn}[n_j^{[stick-free]} - n_j]$$

with the following definitions

$$[7.5] \quad P_a(n_j) = \mathrm{boole}\big[\,|n_j^{[drive]} - n_j| > |n_j^{[drive]} - n_j^{[stick-free]}|^2\,\big] \,|n_j^{[stick-free]} - n_j|\,\mathrm{ran}(a), \qquad \overline{P} \equiv 1 - P,$$

$$[7.6] \quad \left\{ \begin{array}{l} n_j^{[stick-free]} = \mathrm{rd}[\, N_v \dfrac{A_0}{\omega_0 t_{mx}} \sin(2\pi \dfrac{j}{J} + \phi_{eq})\,] \\ n_j^{[drive]} = \mathrm{rd}[\, N_v \cos(2\pi \dfrac{j}{J})\,] \end{array} \right., \qquad N_v \equiv \dfrac{v_{d0}}{\delta v} = 300$$

and for 500 time pixels per period,

$$[7.7] \quad j_{period} = \dfrac{2\pi}{\omega_0 \delta\tau} = 500 \quad \Rightarrow \quad \left\{ \begin{array}{l} M_d^{-1} = \mathrm{rd}[N_v \mu_{dyn} \dfrac{g\delta\tau}{v_{d0}}] = 4 \\ M_{sf}^{-1}(\Delta\mu) = \mathrm{rd}[N_v \dfrac{\Delta\mu}{\mu_{stat}}] \end{array} \right.$$

In Eq. 7.5, we choose the value $aN_v^2 = 6.3$; that is, in a region where appreciable changes have little effect on the results. We checked that the extreme values of the static friction coefficient reproduce the extreme curves: for $\mu_{stat} = \mu_{dyn}$ the slider's curve coincides with the stick-free curve, and for $\mu_{stat} \gg \mu_{dyn}$, it coincides with the driver's curve.

In descriptive words, *Ansatz* 7.4 proposes that, given a velocity $v_{slider}(t_j) \equiv n_j \delta v$, at time $t_{j+1}$ the velocity will either approach the driver ($P$) due to a sticking attempt, or approach the stick-free solution ($\bar{P}$) due to the absence of a sticking attempt. The first term in curly brackets (Eq. 7.4) of the approach to the driver's solution ($P$) contains two Kronecker-delta functions: whenever the previous (sub index $j$) velocity lies on top of the driver curve, so will the next one (sub index $j+1$); else, $M_d^{-1} \text{sgn}[n_{j+1}^{[drive]} - n_j]$ unit velocity steps will be added. The stick-free part ($\bar{P}$) adds $M_{sf}^{-1} \text{sgn}[n_{j+1}^{[stick-free]} - n_j]$ unit velocity steps towards the stick-free solution, for every unsticking attempt. While $M_d^{-1}$ is a fixed number determined by Eq. 7.7, $M_{sf}^{-1}$ is a function of the friction coefficients.

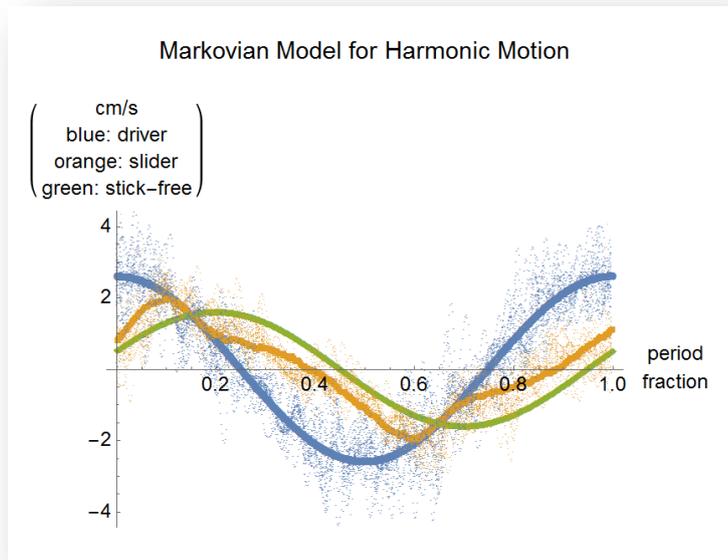

*Figure 7.2: Double-Discrete Markovian simulation of the slider's path for $\mu_{dyn}$ = 32.3%, and $\Delta\mu$ = 4.4‰. Thick curves represent the theoretical and fitted data, and small dots the experimental data (of 18 periods).*

It is remarkable that such a tiny difference in static and dynamic friction coefficients so badly distorts the slider's path: specifically, its near-triangular shape, and the quite impressive phase shift of its response peak from period fraction 20% (of the convolution) to the Markovian value of 10%. The reason for this enormous influence of such a tiny friction difference is due to the extremely low driver acceleration.

A last indication for the usefulness of the here presented model, is that it explains the origin of walking. This is the subject of the last Section.

# 8. Asymmetrically Driven Slider Walk-off

Fig. 8.1 illustrates an asymmetric driver momentum. The integral over a single period vanishes, whence the driver moves asymmetrically around a single point in space. The question of this Section is whether the slider walks off, and if so, in what direction, and why it would walk off at all.

We use the same kick-response as in Section 3 (Eq. 3.3):

[8.1] $\quad K_{kick}(q \mid q_{mx}) \equiv \dfrac{2}{q_{mx}}(1 - \dfrac{q}{q_{mx}})\theta(q)\theta(q_{mx} - q)$

In this section, the driving velocity is not harmonic, but the periodical juxtaposition of two half periods with different frequency (see Fig. 8.1):

[8.2] $\quad w_{drive}(q) = -\theta(q_1)\theta(\tfrac{2}{3} - q_1)\sin \tfrac{3}{2}\pi q_1 + 2\theta(1 - q_1)\theta(q_1 - \tfrac{2}{3})\sin 3\pi q_1$

where we introduced the dimensionless velocity $w$. Eq. 8.2 leads to Fig. 8.1:

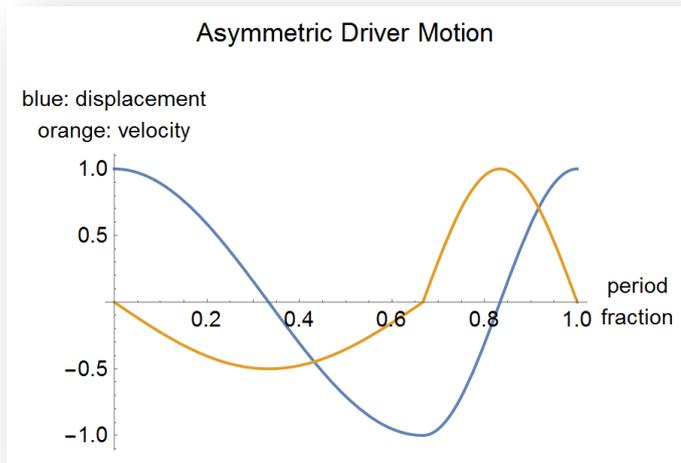

***Figure 8.1: Asymmetric Drive.*** *The high frequency is twice the low one, and occupies one third of a period. Both displayed dimensionless quantities (displacement and velocity) are normalized to unity.*

The velocity (orange line in Fig. 8.1) is the derivative of Eq. 8.2. From now on we stick to the period-fraction unit ($2\pi q \equiv \omega_0 t$) and its modulus ($q_1 \equiv \mathrm{mod}[q,1]$). The convolution reads:

[8.3] $\quad w_{stick-free}(q_t \mid q_{mx}) \equiv \displaystyle\int_0^{q_{mx}} dq\, K_{kick}(q \mid q_{mx}) w_{drive}(q_t - q) = \int_{q_t - q_{mx}}^{q_t} dq\, K_{kick}(q_t - q \mid q_{mx}) w_{drive}(q)$

We will calculate the integral only for the unit window $0 \leq q_t < 1$. Since the stick-free response is periodic, its values outside the mentioned unit window simply follow as

[8.4] $\quad w_{stick-free}(q_t | q_{mx}) = w_{stick-free}(q_{t1} | q_{mx})$

where the modulus-notation applies. Due to the asymmetric nature of the driver velocity, and for $q_{mx} < \tfrac{2}{3}$, The convolution breaks up into four integrals for the following temporal sub-windows:

[8.5] $\quad \begin{cases} \Theta_1 \equiv \theta(q_{t1} - q_{mx})\theta(\tfrac{2}{3} - q_{t1}) \\ \Theta_2 \equiv \theta(q_{t1} - \tfrac{2}{3})\theta(1 - q_{t1}) \\ \Theta_3 \equiv \theta(q_{t1})\theta(\tfrac{2}{3} - q_{t1}) \\ \Theta_4 \equiv \theta[q_{t1} - (\tfrac{2}{3} - q_{mx})]\theta(q_{mx} - q_{t1}) \end{cases}$

The convolution for the unit window thus reads

[8.6] $\quad w_{stick-free}(q_{t1} | q_{mx}) = \sum_{n=1}^{4} \Theta_n(q_{t1} | q_{mx}) J_n(q_{t1} | q_{mx})$

Now define the following general integral

[8.7] $\quad G(q_{t1}, q_{mx} | a, b, c) \equiv \dfrac{1}{q_{mx}} \int_a^b dq\, (1 - \dfrac{q_{t1} - q}{q_{mx}}) \sin c\pi q = \left[ \dfrac{c\pi[(q_{t1} - q - q_{mx})\cos c\pi q] + \sin c\pi q}{(c\pi q_{mx})^2} \right]_{q=a}^{q=b}$

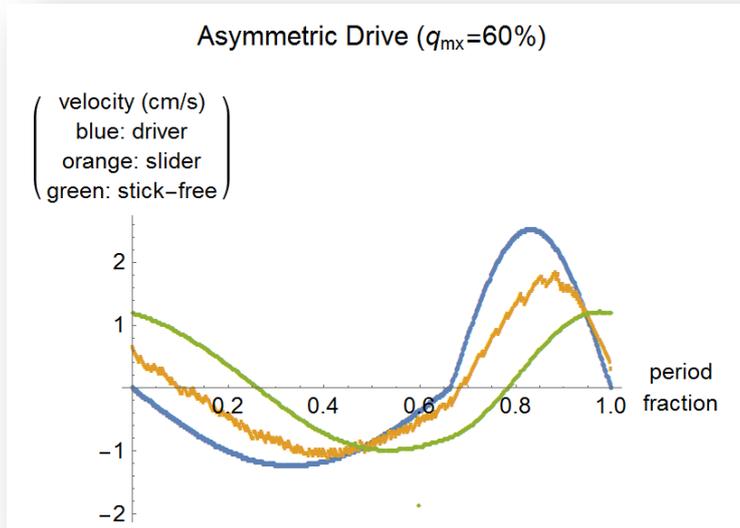

*Figure 8.2: Response to an asymmetric drive.* The integrated surface of the slider's Markovian motion (orange trajectory) is $-11\pm 2$ μm for $q_{mx}$=60%, implying a net shift to the left.

In terms of this integral, the four functions $J_n(q_{t1}, q_{mx})$ of Eq. 8.3 read:

$$[8.8] \begin{cases} J_1(q_{t1}, q_{mx}) \equiv -G(q_{t1}, q_{mx} \mid q_{t1} - q_{mx}, q_t, \tfrac{3}{2}) \\ J_2(q_{t1}, q_{mx}) \equiv -G(q_{t1}, q_{mx} \mid q_{t1} - q_{mx}, \tfrac{2}{3}, \tfrac{3}{2}) + 2G(q_{t1}, q_{mx} \mid \tfrac{2}{3}, q_{t1}, 3) \\ J_3(q_{t1}, q_{mx}) \equiv -G(q_{t1}+1, q_{mx} \mid q_{t1}+1 - q_{mx}, \tfrac{2}{3}, \tfrac{3}{2}) + 2G(q_{t1}+1, q_{mx} \mid \tfrac{2}{3}, 1, 3) - G(q_{t1}, q_{mx} \mid 0, q_{t1}, \tfrac{3}{2}) \\ J_4(q_{t1}, q_{mx}) \equiv 2G(q_{t1}+1, q_{mx} \mid q_{t1}+1 - q_{mx}, 1, 3) - G(q_{t1}, q_{mx} \mid 0, q_{t1}, \tfrac{3}{2}) \end{cases}$$

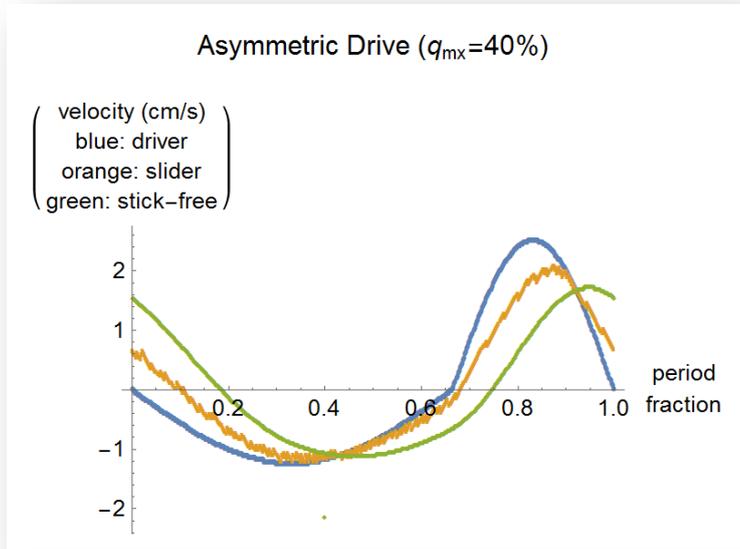

***Figure 8.3: Response to an asymmetric drive.*** *The integrated surface of the slider's Markovian motion (orange trajectory) is −6±2 μm for $q_{mx}$=40%, implying a net shift to the left.*

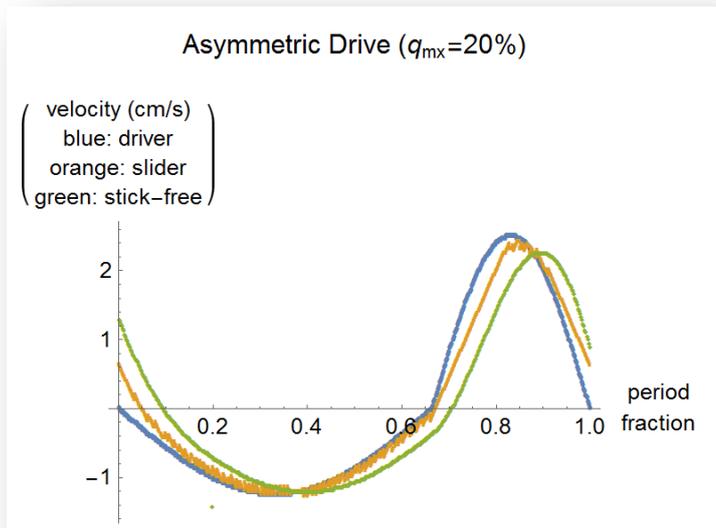

***Figure 8.4: Response to an asymmetric drive.*** *The integrated surface of the slider's Markovian motion (orange trajectory) is −4±2 μm for $q_{mx}$=20%, implying a net shift to the left.*

Combining Eqs. 8.3 through 8.5 yields Figs. 8.2 through 8.5. The only variable parameter was set to the value found in Section 7: $aN_v^2 = 6.3$. The changing variable is the product of harmonic driving frequency and free-flight time, $q_{mx}$, which varies from 20% to 60% over the three figures, which is like varying the effective dynamic friction. In the limit for vanishing $q_{mx}$ (infinite dynamic friction), the response will be indistinguishable from the driver, which implies no walk-off. However, the trend is clear:

(i)      longer kick-responses induce higher walk-off rates;
(ii)      the walk-off sign is opposite to the sign of the driver's acceleration discontinuities;
(iii)      using the same low driver acceleration as in the harmonic experiment, Fig. 8.2 would best describe the walk-off.

# 9. Conclusion

We present experimental data of the motion of a rimmed checker's piece on a horizontal tray, in two specific conditions: on one hand, shaken at 20 Hz by the harmonically moving tray, and on the other, on an equally horizontal, but static, well-polished glass substrate, with a given initial velocity. The latter experiment immediately yields the dynamic friction coefficient. The harmonic experimental results are very sensitive on the static friction coefficient, which is not even a percent higher. From the harmonic data, we deduce that static friction operates via a sticking and un-sticking probability functions, which depend on the slider's velocity and acceleration, thereby generating a Markovian process. The following step towards a solid Markovian friction model is its generalization to describing the slider's behavior as a function of driver acceleration. This fundamentally requires a pile of trustworthy experimental data, as the Markovian progression algorithm is easily adapted.

# Appendix: Smoothed Data and Noise Calculation

1. The Fourier filter used to smooth the raw data (see Fig. 2.1) is

[A1] $\quad F_x(\nu) = e^{-(\frac{\nu}{\nu_0})^2}(1 - e^{-(\frac{\nu}{\nu_1})^4})$

with $\nu_0 = 400\,\delta\nu$, $\nu_1 = 20\,\delta\nu$, and the frequency $\nu$ extending from to $-4500$ Hz to $+4500$ Hz. The frequency spacing is

[A2] $\quad \delta\nu = \dfrac{1}{T} = \dfrac{\nu_0}{18} = 1.11\,Hz$

The dimensionless number 18 represents the number of periods covered by the experimental data $x(t)$. The Gaussian exponent rejects all frequencies beyond $800\,\delta\nu$. The second factor rejects all frequencies below 30 $\delta\nu$, such that the back-transformed result has practically no offset. Note that the second exponent has a fourth power, not a second; and that the Fourier Transform is discrete, in spite of the continuous appearance of Eq. A1.

2. The frequency filter, shown in red in Fig. A2, is the basis of our noise calculation.

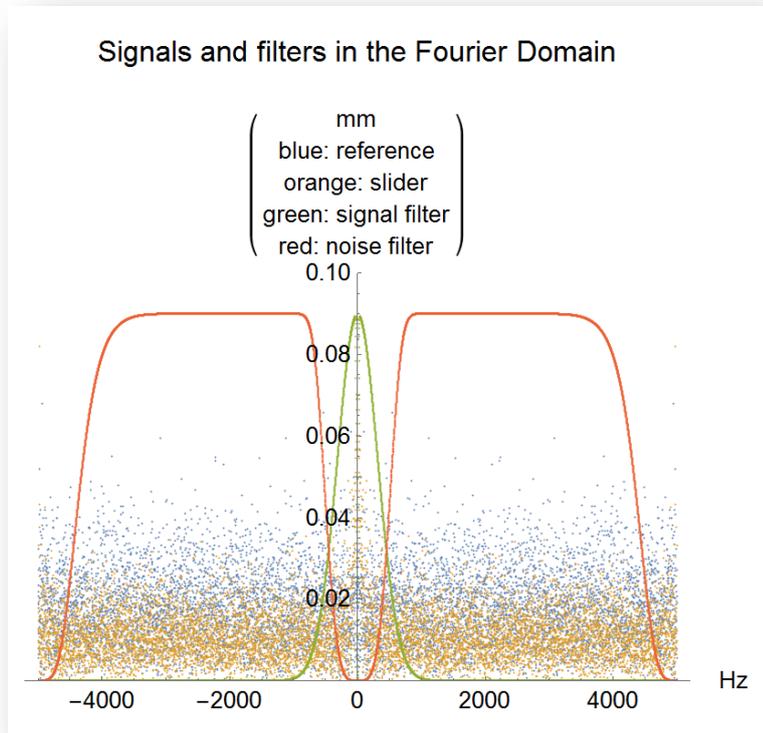

*Figure A1: Signals and filters in the Fourier Domain. The mm scale for the vertical axis only applies to the signals (dots). Around zero frequency the signals reach up to 20 mm, possibly due to an offset. The*

green filter is the signal filter (Eq. A1), yielding the smoothed data presented in Fig. 2.1. The red curve is the noise filter (Eq. A3), blocking both signal and digitizing errors.

The noise filter is

[A3] $\quad F_{N[x]}(v) = e^{-(\frac{v}{v_2})^4} (1 - e^{-(\frac{v}{v_3})^{20}})$

with $v_2 = 4000\, \delta v$, $v_3 = 500\, \delta v$, and is shown in red in Fig. A1. This filter blocks all digitizing errors as well as all signal frequencies, thus keeping only the experimental noise. The spectrum was then back-transformed to the time-domain, spanning 18 driving periods (a single period corresponding to 50 ms). Fig. A2 shows the average of the 18-period time-domain.

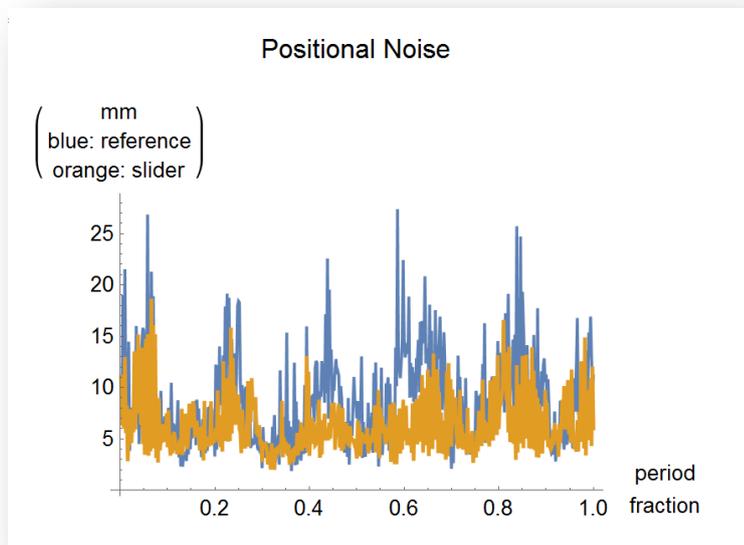

*Figure A2: Noise on the Position Data. Note that the horizontal axis represents time, not frequency.*